\documentclass[conference]{IEEEtran}
\usepackage{graphicx}
\usepackage[fleqn]{amsmath}
\usepackage[justification=centering]{caption}  
\usepackage{cite}                              
\usepackage[numbers,sort&compress]{natbib}     
\usepackage{multirow}
\def\BibTeX{{\rm B\kern-.05em{\sc i\kern-.025em b}\kern-.08em T\kern-.1667em\lower.7ex\hbox{E}\kern-.125emX}}

\ifCLASSINFOpdf

\else

\fi

\begin{document}

\title{A Novel Secure Authentication Scheme for Heterogeneous Internet of Things}
\author{\IEEEauthorblockN{Jingwei Liu\IEEEauthorrefmark{1},
Ailian Ren\IEEEauthorrefmark{1},
Lihuan Zhang\IEEEauthorrefmark{1},
Rong Sun\IEEEauthorrefmark{1}, and
Mohsen Guizani\IEEEauthorrefmark{2}
}
\IEEEauthorblockA{\IEEEauthorrefmark{1}State Key Lab of ISN, Xidian University, Xi'an, 710071, China.\\ Email: \{jwliu, rsun\}@mail.xidian.edu.cn, 979863819@qq.com, zhanglihuan678@163.com}
\IEEEauthorblockA{\IEEEauthorrefmark{2}Department of Electrical and Computer Engineering, University of ldaho, Mosocow, ldaho, USA.\\ Email: mguizani@ieee.org}
}
\maketitle

\begin{abstract}

Today, Internet of Things (IoT) technology is being increasingly popular which is applied in a wide range of industry sectors such as healthcare, transportation and some critical infrastructures. With the widespread applications of IoT technology, people's lives have changed dramatically. Due to its capabilities of sensitive data-aware, information collection, communication and processing, it raises security and privacy concerns. Moreover, a malicious attacker may impersonate a legitimate user, which may cause security threat and violation privacy. In allusion to the above problems, we propose a novel and lightweight anonymous authentication and key agreement scheme for heterogeneous IoT, which is innovatively designed to shift between the public key infrastructure (PKI) and certificateless cryptography (CLC) environment. The proposed scheme not only achieves secure communication among the legal authorized users, but also possesses more attributes with user anonymity, non-repudiation and key agreement fairness. Through the security analysis, it is proved that the proposed scheme can resist replay attacks and denial of service (DOS) attacks. Finally, the performance evaluation demonstrates that our scheme is more lightweight and innovative.

\end{abstract}

\IEEEpeerreviewmaketitle

\section{Introduction}
With the rapid development of modern smart technologies, Internet of things (IoT) has caught much attention from industry and IT community in terms of networking and communication aspects \cite{Yaqoob2017Internet}. In the future, IoT communication scenarios will be a combination of heterogeneous access technologies and services, which enables users to be exposed to a diverse network environment. The heterogeneity of IoT determines that information can flow among multiple transmission networks with different structures, providing various services on a common network platform. Like in all other communication and computer networks, security issues are always significantly important in the development of heterogeneous IoT (HIoT). In addition, key agreement and authentication mechanism play indispensable roles in the aspects of protecting user privacy and data security for HIoT scenario.

\begin{figure}[!t]
\centering
\includegraphics[width=8.5cm]{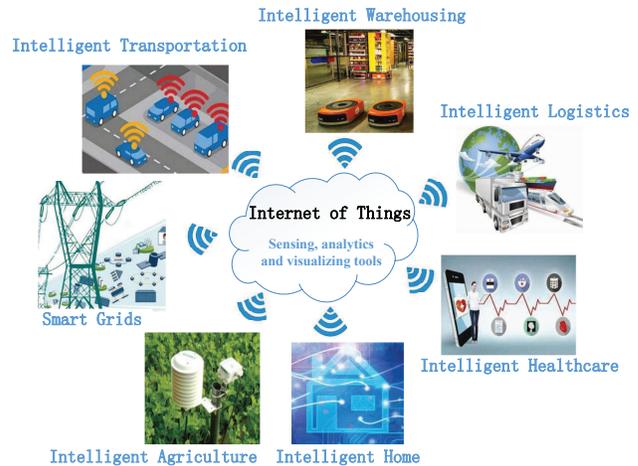}\\
\caption{Application Scenario of IoT}
\label{Figure 1}
\end{figure}

IoT is a popular notion \cite{Wu2017A} that has been widely used in industries such as healthcare and some critical infrastructures, as shown in Fig.\ref{Figure 1}. Meanwhile, the diversity of IoT applications and heterogeneity of IoT communication infrastructures also have led to lots of security challenges \cite{Du2008Security,Du2008Defending,Du2009A,Xiao2007A,Du2007An,Xiao2007Internet,Du2005Designing,Du2006Adaptive,Du2004QoS,Mandala2006}, exposing some threats of malicious attacks, data interception, user privacy leaking, unauthorized access, etc. These security flaws seriously affect HIoT's development and application. Therefore, security and privacy become essential in HIoT.

In 2010, the EU Commission \cite{Sundmaeker2010Vision} identified security and privacy as a major IoT research challenge. Many researches \cite{Turkanovic2014A,Xu2017hybrid,Wei2017A,Farash2016An} focused on secure communication and privacy preservation for HIoT. Moreover, some other researches \cite{Feng2017A,Zhang2015Communication,Kim2012An,Wang2015Preserving} put forward countermeasures that were targeted at specific types of attacks. In 2017, Feng et al. \cite{Feng2017A} presented a replay-attack resistant authentication scheme, based on an improved challenge-response mechanism instead of the timestamp mechanism. In \cite{Zhang2015Communication}, a lightweight defensive algorithm for distributed denial of service attacks (DDOS) was proposed for IoT environment, which could protect the sensor nodes from the attacks of malicious requests effectively. In order to preserve user privacy, Kim-Kim's scheme \cite{Kim2012An} adopted the one-time pseudonym identities synchronization mechanism that could maintain identities consistency between users and the server. However, the scheme was vulnerable to the de-synchronization attack. In 2015, Wang et al. \cite{Wang2015Preserving} improved the Kim-Kim's scheme, achieving superior privacy preservation.

Authentication and key agreement are the core technologies and the foundation of other security mechanisms. It enables legal authorized users to establish a reliable relationship between each other in HIoT. In 2009, a key establishment and authentication scheme based on combined public key (CPK) algorithm was proposed for the heterogeneous network, and it was proved to be efficient in terms of the mutual authentication \cite{Li2009Key}. In 2013, Chu et al. \cite{Chu2013An} proposed an identity authentication scheme based on elliptic curve cryptographic (ECC), which innovatively used the encryption algorithm of public-private key pair to satisfy the security requirements of heterogeneous network. In 2016, Amin et al. \cite{Amin2016Design} came up with a three-factor authenticated key agreement scheme for IoT and claimed that their scheme was secure. However, Arasteh et al. \cite{Arasteh2016A} showed that the scheme of Amin et al. was prone to replay attacks and DOS attacks.

Moreover, due to the resource-constrained nature of IoT devices, the secure schemes should be lightweight. In allusion to this requirement, some schemes \cite{Mahmood2017Lightweight,Sarvabhatla2014A} were proposed to reduce the computation burden of participants. In 2016, Iqbal and Bayoumi \cite{Iqbal2016A} proposed a novel authentication and key agreement scheme, which offloaded the heavy cryptographic functions of resource-constrained sensors to the trusted neighboring sensors. In \cite{Alkuhlani2017Lightweight}, a secure and lightweight mutual authentication and key agreement scheme was presented, in which the cryptographic functions were proved to be computationally lightweight and resist some known attacks. Although these schemes were lightweight in the IoT environment, they did not prove that they were applicable to heterogeneous IoT. In \cite{Hou2008CPK}, Hou et al. proposed a secure and lightweight authentication and key agreement scheme based on CPK and ECC in HIoT environment. Unfortunately, this scheme used signature-to-encryption in a time-consuming way.

In this paper, we propose a novel and lightweight anonymous authentication and key agreement scheme for heterogeneous IoT, which is based on a signcryption algorithm between PKI and CLC environment. It provides more features of user anonymity, non-repudiation, key agreement fairness and lightweight. In addition, our scheme can be proved to resist replay attacks and DOS attacks. Therefore, it has wider application prospect in HIoT environment.

The rest of this paper is organized as follows. In Section II, we discuss the secure HIoT's system model. Section III demonstrates the proposed mutual authentication and key agreement scheme for HIoT. Sections IV and V present informal security and performance analysis respectively. At last, the conclusion is described in section VI.

\section{Preliminaries}
In this section, there is a brief description of the HIoT's system model and security assumptions.

~\\\textbf{A. System model}

\begin{figure}[!t]
\centering
\includegraphics[width=8.5cm]{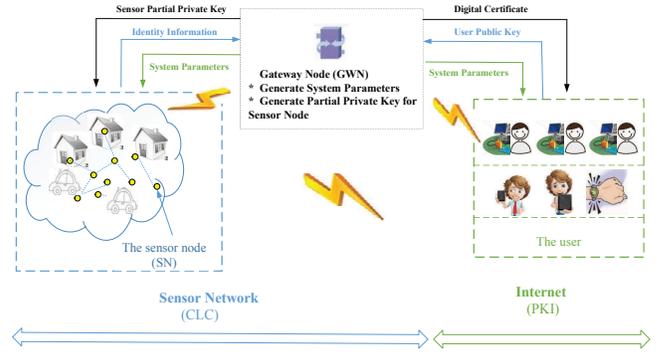}\\
\caption{A secure system model for HIoT}
\label{Figure 2}
\end{figure}

A typical scenario model of HIoT is illustrated in Fig. \ref{Figure 2}. It mainly consists of a gateway node (GWN), a user and a sensor node (SN).

The GWN is a trust third party that can distribute partial private key to the SN and the digital certificate to the user respectively. The SN in CLC is in charge of gathering data from environments and forwarding the data to the user in PKI via a secure channel.

~\\\textbf{B. Security assumptions}

The security of the proposed scheme depends on the hardness of the following problems:

$G_{1}$ is a cyclic additive group, and $G_{2}$ is a cyclic multiplicative group. A large prime $q$ is the order of $G_{1}, G_{2}$. $e(.,.)$ is a bilinear map $G_{1} \times G_{1} \rightarrow G_{2}$. $P$ is a generator of $G_{1}$ and $g=e(P,P)$.

~\\\textbf{Definition 1 (CDHP)}. Defining Computational Diffie-Hellman Problem (CDHP) is to compute $abP \in G_{1}$ when given $(P,aP,bP)\in G_{1}$.

~\\\textbf{Definition 2 (ECDLP)}. Defining Elliptic Curve Discrete Logarithm Problem (ECDLP) is to compute the integer $a \in Z_{q}^*$ when given $(P,aP) \in G_{1}$.

\section{The Proposed Scheme}

\begin{figure*}[!t]
\centering
\includegraphics[width=\textwidth]{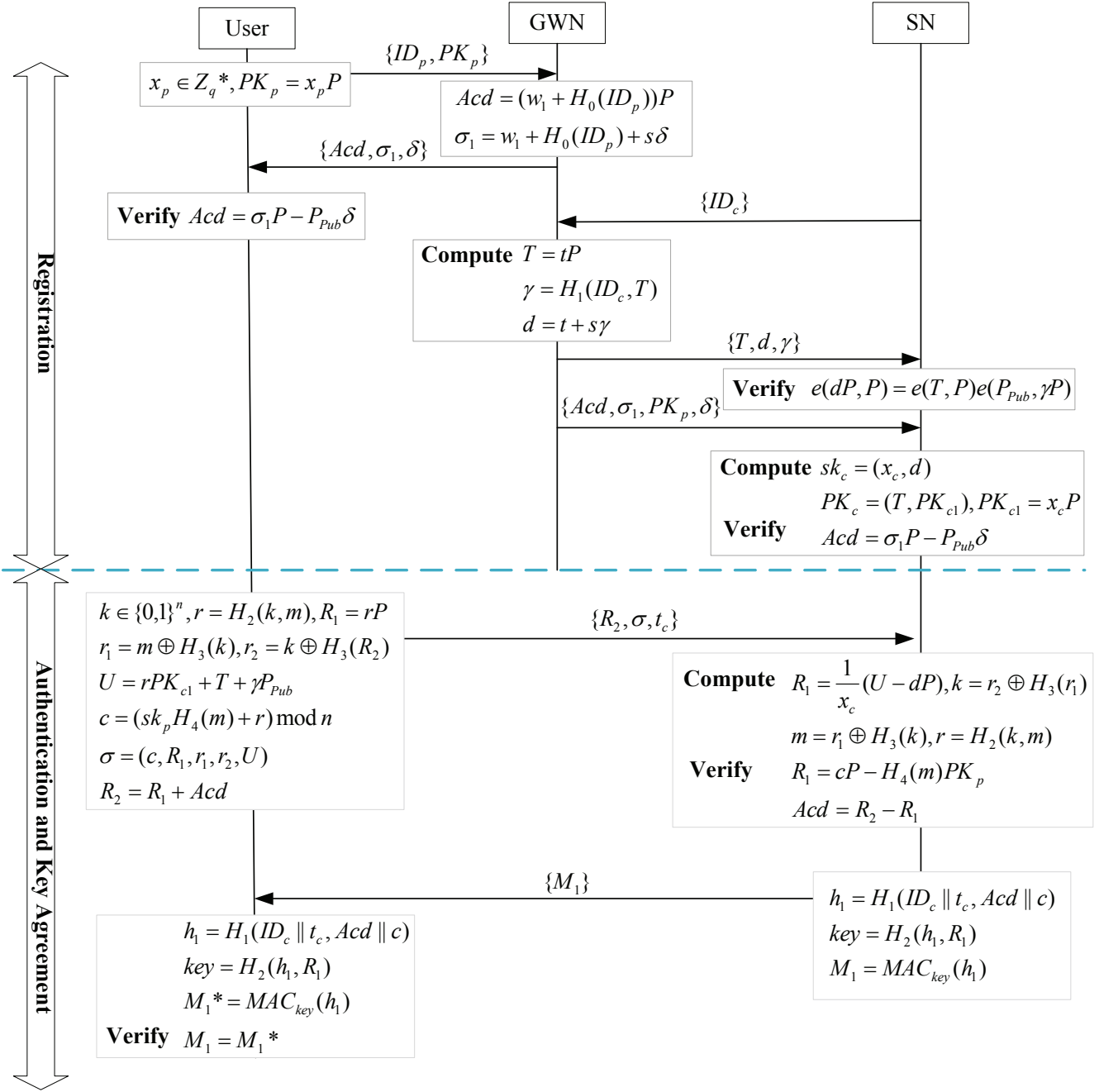}\\
\caption{Simplified overall architecture of our scheme}
\label{Figure 3}
\end{figure*}

In this section, we propose a novel anonymous authentication and key agreement scheme based on a signcryption algorithm for HIoT, as shown in Fig. \ref{Figure 3}. The proposed scheme comprises three phases: system initialization, system registration, system authentication and key agreement phase.

~\\\textbf{A. System initialization phase.}

1) The GWN selects the main private key $s\in Z_{q}^{*}$ randomly, and calculates the public key $P_{pub}=sP$. Let $l$ be the security parameter of the system and $ID=\{0,1\} ^*$ be an identity space.

2) The GWN defines five secure cryptographic hash functions: $H_{0}:\{0,1\} ^*\rightarrow Z_{q}^*$, $H_{1}:\{0,1\}^*\times G_{1}\rightarrow Z_{q}^*$, $H_{3}:\{0,1\}^n\times G_{1}\rightarrow Z_{q}^*$, $H_{4}:\{0,1\}^n\rightarrow Z_{q}^*$. Then, it publishes $parameter\{G_{1},G_{2},P,P_{pub},l,g,H_{0},H_{1},H_{2},H_{3},H_{4}\}$ and keeps $s$.

~\\\textbf{B. System registration phase.}

This phase is divided into two steps: registration between the user and the GWN, registration between the SN and the GWN.

\textbf{Step 1:} Registration between the user in PKI environment and the GWN.

1) The user runs PKI-Key-Gen algorithm:

\begin{itemize}
\item  Select $x_{p}\in Z_{q}^{*}$ randomly as the private key $sk_{p}=x_{p}$;
\item  Compute $P_{pub}=x_{p}P$ as its public key;
\item  Send the message $\{ID_{p},PK_{p}\}$ to the GWN through a secure channel.
\end{itemize}

2) The GWN firstly computes the account information $Acd=(w_{1}+H_{0}(ID_{p}))P$ and the signature information $\sigma_{1}=w_{1}+H_{0}(ID_{p})+s\delta$, where $w_{1}\in Z_{q}^{*}$ and $\delta=H_{1}(ID_{p},PK_{p})$. Then, it returns the identity account management information $\{Acd,\sigma_{1},\delta\}$ to the user.

3) The user checks if the equation $Acd=\sigma_{1}P-P_{pub}\delta$ holds or not.

\begin{itemize}
\item  If it holds, the user stores $\{Acd,\sigma_{1},\delta\}$, and regards $Acd$ as its identifier, which indicates the registration between the GWN and the user is successful;
\item  Otherwise, the user aborts.
\end{itemize}

~\\\textbf{Step 2:} Registration between the SN in CLC environment and the GWN.

1) The SN sends its identity $ID_{c}$ to the GWN.

2) The GWN runs CLC-Partial-Private-Key-Gene algorithm:

\begin{itemize}
\item  Select $t\in Z_{q}^{*}$ randomly, compute $T=tP$, $\gamma=H_{1}(ID_{c},T)$ and obtain part of the private key $d=t+s\gamma$;
\item  Return $\{T,d,\gamma\}$ to the SN through a secure channel.
\end{itemize}

3) The SN checks if $e(dP,P)=e(T,P)e(P_{pub},\gamma P)$ holds or not.

\begin{itemize}
\item  If it holds, the GWN is legal. Then, the SN selects $x_{c}\in Z_{q}^{*}$ and obtains the intact privacy key $sk_{c}=\{x_{c},d\}$;
\item  Otherwise, the SN aborts.
\end{itemize}

4) The SN computes $PK_{c1}=x_{c}P$ and sets $PK_{c}=\{T,PK_{c1},\gamma\}$ as its intact public key. Then, it checks if $Acd=\sigma_{1}P-P_{pub}\delta$.

\begin{itemize}
\item  If it holds, the SN stores $\{Acd,\sigma_{1},PK_{p},\delta\}$, and regards $Acd$ as legal user's identifier;
\item  Otherwise, the SN aborts.
\end{itemize}

~\\\textbf{C. System authentication and key agreement phase.}

1) The user runs the PKI-to-CLC heterogeneous signcryption algorithm:

\begin{itemize}
\item Select a random number $k\in \{0,1\}^{n}$;

\item Compute $r=H_{2}(k,m)$, $R_{1}=rP$, $r_{1}=m\oplus H_{3}(k)$, $r_{2}=k\oplus H_{3}(r_{1})$;

\item Compute $U=rPK_{c1}+T+\gamma P_{pub}$;

\item Compute $c=(sk_{p}H_{4}(m)+r)\pmod n$;

\item Obtain the ciphertext message $\sigma=\{c,R_{1},r_{1},r_{2},U\}$;

\item Calculate the account protection information $R_{2}=R_{1}+Acd$;

\item Send a service request message $\{R_{2},\sigma,t_{c}\}$ to the SN.
\end{itemize}

2) Then, the SN runs the PKI-to-CLC heterogeneous unsigncryption algorithm:

\begin{itemize}
\item Check if $|t_{c}-t_{1}|<\Delta t$, where $t_{1}$ is the current timestamp, and $\Delta t$ is the allowed transmission delay. If it holds, the message is fresh. The SN can carry on the subsequent steps. Otherwise, it terminates the authentication.

\item Compute $R_{1}=\frac{1}{x_{c}}(U-dP)$;

\item Compute $k=r_{2}\oplus H_{3}(r_{1})$, $m=r_{1}\oplus H_{3}(k)$, $r=H_{2}(k,m)$;

\item If $R_{1}=cP-H_{4}(m)PK_{p}$, the SN receives $\sigma$. Otherwise, it returns $\bot$ for rejection;

\item Check if $Acd=R_{2}-R_{1}$. If it does, the SN computes its own message digest $h_{1}=H_{1}(ID_{c}||t_{c},Acd||c)$. Otherwise, the SN aborts;

\item Calculate the session key $key=H_{2}(h_{1},R_{1})$;

\item Calculate the message authentication code $M_{1}=MAC_{key}(h_{1})$ and send $M_{1}$ to the user.
\end{itemize}

3) Finally, the user does the following verification:

\begin{itemize}
\item Calculate the message digest $h_{1}=H_{1}(ID_{c}||t_{c},Acd||c)$;

\item Calculate the session key $key=H_{2}(h_{1},R_{1})$;

\item Generate a new message authentication code $ M_{1}^{*}=MAC_{key}(h_{1})$. And it checks if $M_{1}=M_{1}^{*}$. If it does, the authentication is successful. Otherwise, it is failed.
\end{itemize}

If mutual authentication is successful, the user and the SN could generate the same session key respectively. Therefore, they can communicate with each other securely in the future.

\section{The security analysis}

\subsection{Mutual authentication}
The scheme realizes mutual heterogeneous authentication between SNs and users.
When a user starts the authentication phase with a SN, the SN has to verify the identity of the user. Thereby, it needs to verify the user's legality by running the authentication phase of Section III. After verifying user's legality, the user starts authenticating the SN. The user can also verify the SN's legality by running the authentication phase of Section III. When the SN and the user are both proven to be legal, they also have completed the key agreement phase.

\subsection{Key agreement}
Only the user and the SN can get the session key $key=H_{2}(h_{1},R_{1})$. The user does not send the account information $Acd$ to the SN directly. $Acd$ is hidden in the account protection information $R_{2}=R_{1}+Acd$. If the private key of the SN is unavailable, anyone else can not obtain $Acd$ and $h_{1}$. In other word, it is impossible for others to get the session key. Therefore, the session key is secure in our scheme.

\subsection{Anonymity}
Our scheme ensures the user's anonymity via the masked identity $\{ID_{p},PK_{p}\}$. In system registration phase, the user sends the registration request message $\{ID_{p},PK_{p}\}$ to the GWN. The GWN creates an account information $Acd$ for the user. Here, $Acd$ is not the real identity of the user. Namely, it is only the user's identifier. When the user sends a service request message to the SN, the SN sends the corresponding service to the user with $Acd$, but it does not know the real identity of the user. Because the SN can not derive the user's identity information from $Acd$. So our scheme achieves the user's identity anonymity in system registration phase.

In system authentication phase, $Acd$ is not transmitted in plaintext and is hidden by the account protection information $R_{2}=R_{1}+Acd$. Only the user and the SN can get $Acd$. Even the GWN can not get the user's real identity. Therefore, our scheme can provide the user's identity anonymity.

\subsection{Non-repudiation}
The user sends the service request message $\{R_{2},\sigma,t_{c}\}$ to the SN, but in this process others can not forge the ciphertext message $\sigma$ without obtaining the user's intact private key due to the hardness of the $CDHP$ problem under current conditions. Thereby, $\sigma$ can only be generated by the user. When the authentication is successful, the SN will provide the corresponding services for the legal user. The user can not deny sending the service request messages to the SN. Similarly, anyone else is unable to impersonate the SN due to the lack of the SN's private key. Therefore, the SN can not deny either that it had received the user's service request messages or that it had provided the corresponding services for the user.

\subsection{Key agreement fairness}
The SN calculates the session key $key=H_{2}(h_{1},R_{1})$ and the message authentication code $M_{1}$. Then it sends $M_{1}$ to the user. After receiving $M_{1}$, the user calculates the message digest $h_{1}=H_{1}(ID_{c}||t_{c},Acd||c)$ in order to get the session key $key=H_{2}(h_{1},R_{1})$. It computes the message authentication code $M_{1}^{*}$. The SN and the user can get the session key and the message authentication code equally, and one participant does not have more privilege than the other. Therefore, the communication participants are in an equal position after the key agreement is completed. According to the above, fairness is ensured in our scheme.

\subsection{Anti-replay attacks}
The SN is unable to identify the validity of the message from the user, because it does not know if the message had been received by itself. The attackers usually utilize the drawback to initiate a replay attack on the SN. Our scheme avoids this drawback by involving a timestamp to the user's service request message effectively. After receiving the service request message, the SN can check the freshness of the message based on the judgment of the timestamp to identify if the message could be accepted. Therefore, our scheme can resist replay attacks.

\subsection{Denial of service attacks}
In system registration phase, the GWN sends the account information $Acd$ to the user and the SN respectively through a secure channel. When the user initiates a service request for the SN, the SN utilizes the account protection information $R_{2}$ and own private key to compute $Acd$. Then, it verifies if $Acd$ is equal to the received $Acd$, so as to determine if it should accept the service request message from the user. Before generating a bogus service request, the attackers must calculate $Acd$ correctly ahead. However, they can not get the correct $Acd=(w_{1}+H_{0}(ID_{p}))P$ without the random number $w_{1}$. Our scheme exploits $Acd$ to prevent attackers from abusing system resources to send a lot of invalid service request messages. Therefore, our scheme can resist denial of service attacks successfully.

\section{The performance evaluation}
For quantitative analysis of our scheme, we use Ubuntu OS as the experimental platform to simulate the total running time. Let P denote the bilinear pairing operation, M denote the point multiplication operation in $G_{1}$, E denote the exponential operation in $G_{2}$, and H denote the hash operation separately.

\begin{figure*}[!t]
\centering
\includegraphics[width=9.5cm]{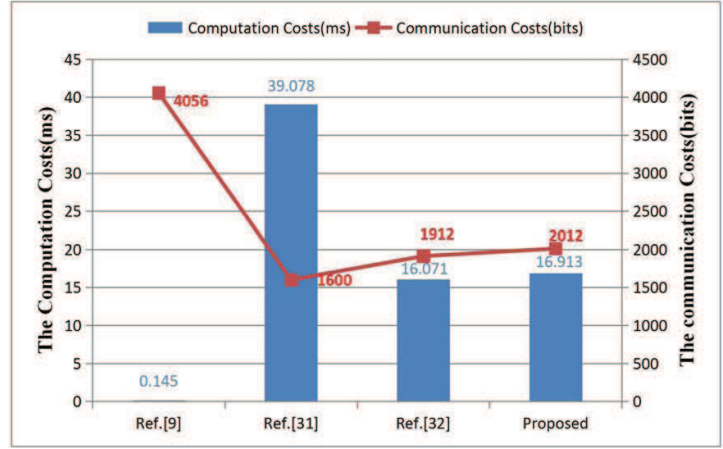}\\
\caption{The total consumption time and communication costs of each scheme}
\label{Figure 5}
\end{figure*}

TABLE \ref{Table I} demonstrates the performance comparison between the proposed scheme and other related schemes. Note that SC denotes symmetric cryptography. Fig. \ref{Figure 5} shows the total computation and communication costs of each scheme.

\begin{table*}[!t]
\centering  
\caption{Performance comparison}
\label{Table I}
\begin{tabular}{|c|c|c|c|c|c|c|c|} 
\hline
\multirow{2}{*}{}  &\multicolumn{4}{|c|}{Computation costs}  &\multirow{2}{*}{} &\multirow{2}{*}{}  &\multirow{2}{*}{} \\
\cline{2-5}
Schemes                             &User(or Alice)       &GWN(or PKG)        &SN(or Bob)         &Total time(ms)      &Communication costs        &Signcryption-based          &Domains\\
\hline
Ref. \cite{Farash2016An}                &19H                  &17H                &9H                 &0.145                  &4056bits                    &No                      &SC \\
\cline{1-8}
Ref. \cite{Wang2017An}             &5P+3E+9M+6H              &H+M             &5P+2E+6M+6H            &39.078                 &1600bits                    &Yes                   &PKI-IBC\\
\cline{1-8}
Ref. \cite{Luo2018Secure}           &P+E+3M+2H               &4M+4H             &4P+E+2H              &16.071                 &1912bits                    &Yes                   &IBC-CLC\\
\cline{1-8}
Proposed                             &3M+4H                  &2M+2H             &3P+4M+6H             &16.913                 &2012bits                    &Yes                   &PKI-CLC\\
\hline
\end{tabular}
\end{table*}

From TABLE \ref{Table I}, we can clearly find that the scheme \cite{Farash2016An} does not involve signcryption algorithm, so it causes higher communication costs. Although the scheme \cite{Wang2017An} has a lower communication overhead, it is inefficient with higher computation complexity. Furthermore, the scheme in \cite{Luo2018Secure} has only proved that it satisfies confidentiality and unforgeability. Besides, it has the key escrow issue, causing a lot of storage space to be occupied for the resource-constrained sensor node side. For all above, our scheme is more applicable for heterogeneous IoT.

\section{Conclusion}
In recent time, the security and privacy issues of HIoT have drawn much attention from all walks of life. In order to solve these problems for HIoT, this paper proposes an anonymous mutual authentication and key agreement scheme based on a proper signcryption algorithm. The proposed scheme not only possesses the features of anonymity, non-repudiation, key agreement fairness, but also can resist replay attacks and DOS attacks. Additionally, the scheme is highly lightweight with lower computation and communication overhead. What's more, it is innovatively designed to shift between the PKI and CLC environment. As a consequence, our scheme has a better scalability to be more applicable for heterogeneous IoT.

\ifCLASSOPTIONcaptionsoff
  \newpage
\fi

\bibliographystyle{IEEEtran}
\bibliography{ms}

\begin{thebibliography}{10}
\providecommand{\url}[1]{#1}
\csname url@samestyle\endcsname
\providecommand{\newblock}{\relax}
\providecommand{\bibinfo}[2]{#2}
\providecommand{\BIBentrySTDinterwordspacing}{\spaceskip=0pt\relax}
\providecommand{\BIBentryALTinterwordstretchfactor}{4}
\providecommand{\BIBentryALTinterwordspacing}{\spaceskip=\fontdimen2\font plus
\BIBentryALTinterwordstretchfactor\fontdimen3\font minus
  \fontdimen4\font\relax}
\providecommand{\BIBforeignlanguage}[2]{{%
\expandafter\ifx\csname l@#1\endcsname\relax
\typeout{** WARNING: IEEEtran.bst: No hyphenation pattern has been}%
\typeout{** loaded for the language `#1'. Using the pattern for}%
\typeout{** the default language instead.}%
\else
\language=\csname l@#1\endcsname
\fi
#2}}
\providecommand{\BIBdecl}{\relax}
\BIBdecl

\bibitem{Yaqoob2017Internet}
I.~Yaqoob, E.~Ahmed, I.~A.~T. Hashem, A.~I.~A. Ahmed, A.~Gani, M.~Imran, and
  M.~Guizani, ``Internet of things architecture: Recent advances, taxonomy,
  requirements, and open challenges,'' \emph{IEEE Wireless Communications
  Magazine}, vol.~24, no.~3, pp. 10--16, 2017.

\bibitem{Wu2017A}
F.~Wu, L.~Xu, S.~Kumari, and X.~Li, ``A privacy-preserving and provable user
  authentication scheme for wireless sensor networks based on internet of
  things security,'' \emph{Journal of Ambient Intelligence and Humanized
  Computing}, vol.~8, no.~1, pp. 101--116, 2017.

\bibitem{Du2008Security}
X.~Du and H.~H. Chen, ``Security in wireless sensor networks,'' \emph{IEEE
  Wireless Communications Magazine}, vol.~15, no.~4, pp. 60--66, 2008.

\bibitem{Du2008Defending}
X.~Du, M.~Guizani, Y.~Xiao, and H.~H. Chen, ``Defending dos attacks on
  broadcast authentication in wireless sensor networks,'' in \emph{Proc. of
  IEEE International Conference on Communications}, 2008, pp. 1653--1657.

\bibitem{Du2009A}
------, ``A routing-driven elliptic curve cryptography based key management
  scheme for heterogeneous sensor networks,'' \emph{IEEE Transactions on
  Wireless Communications}, vol.~8, no.~3, pp. 1223--1229, 2009.

\bibitem{Xiao2007A}
Y.~Xiao, V.~K. Rayi, B.~Sun, X.~Du, F.~Hu, and M.~Galloway, ``A survey of key
  management schemes in wireless sensor networks,'' \emph{Computer
  Communications}, vol.~30, no. 11-12, pp. 2314--2341, 2007.

\bibitem{Du2007An}
X.~Du, Y.~Xiao, M.~Guizani, and H.~H. Chen, ``An effective key management
  scheme for heterogeneous sensor networks,'' \emph{Ad Hoc Networks}, vol.~5,
  no.~1, pp. 24--34, 2007.

\bibitem{Xiao2007Internet}
Y.~Xiao, X.~Du, J.~Zhang, F.~Hu, and S.~Guizani, ``Internet protocol television
  (iptv): the killer application for the next generation internet,'' \emph{IEEE
  Communications Magazine}, vol.~45, no.~11, pp. 126--134, 2007.

\bibitem{Du2005Designing}
X.~Du and F.~Lin, ``Designing efficient routing protocol for heterogeneous
  sensor networks,'' in \emph{Proc. of the 24th IEEE International Performance,
  Computing, and Communications Conference (IPCCC)}, 2005.

\bibitem{Du2006Adaptive}
X.~Du and D.~Wu, ``Adaptive cell-relay routing protocol for mobile ad hoc
  networks,'' \emph{IEEE Transactions on Vehicular Technology}, vol.~55, no.~1,
  pp. 270--277, 2006.

\bibitem{Du2004QoS}
X.~Du, ``Qos routing based on multi-class nodes for mobile ad hoc networks,''
  \emph{Ad Hoc Networks}, vol.~2, no.~3, pp. 241--254, 2004.

\bibitem{Mandala2006}
D.~Mandala, F.~Dai, X.~Du, and C.~You, ``Load balance and energy efficient data
  gathering in wireless sensor networks,'' in \emph{Proc. of the First IEEE
  International Workshop on Intelligent Systems Techniques for Wireless Sensor
  Networks, in conjunction with IEEE MASS'06}, 2006.

\bibitem{Sundmaeker2010Vision}
H.~Sundmaeker, P.~Guillemin, P.~Friess, and S.~Woelffl\'e, ``Vision and
  challenges for realizing the internet of things,'' \emph{International
  Journal of Systematic {\&} Evolutionary Microbiology}, vol.~73, no.~1, pp.
  55--70, 2010.

\bibitem{Turkanovic2014A}
M.~Turkanovi\'c, B.~Brumen, and M.~H\"olbl, ``A novel user authentication and
  key agreement scheme for heterogeneous ad hoc wireless sensor networks, based
  on the internet of things notion,'' \emph{Ad Hoc Netwoeks}, vol.~20, no.~2,
  pp. 96--112, 2014.

\bibitem{Xu2017hybrid}
Y.~Xu and F.~Liu, ``Hybrid key management scheme for preventing man-in-middle
  attack in heterogeneous sensor networks,'' in \emph{Proc. of 2017 3rd IEEE
  International Conference on Computer and Communications (ICCC)}, 2017, pp.
  1421--1425.

\bibitem{Wei2017A}
F.~Wei, P.~Vijayakumar, J.~Shen, R.~Zhang, and L.~Li, ``A provably secure
  password-based anonymous authentication scheme for wireless body area
  networks,'' \emph{Computers {\&} Electrical Engineering}, vol.~65, pp.
  322--331, 2018.

\bibitem{Farash2016An}
M.~S. Farash, M.~Turkanovi\'c, S.~Kumari, and M.~H\"olbl, ``An efficient user
  authentication and key agreement scheme for heterogeneous wireless sensor
  network tailored for the internet of things environment,'' \emph{Ad Hoc
  Networks}, vol.~36, no.~P1, pp. 152--176, 2016.

\bibitem{Feng2017A}
Y.~Feng, W.~Wang, Y.~Weng, and H.~Zhang, ``A replay-attack resistant
  authentication scheme for the internet of things,'' in \emph{Proc. of 2017
  IEEE International Conference on Computational Science and Engineering (CSE)
  and IEEE International Conference on Embedded and Ubiquitous Computing
  (EUC)}, 2017, pp. 541--547.

\bibitem{Zhang2015Communication}
C.~Zhang and R.~Green, ``Communication security in internet of thing:
  preventive measure and avoid ddos attack over iot network,'' in \emph{Proc.
  of the 18th Symposium on Communications {\&} Networking}, 2015, pp. 8--15.

\bibitem{Kim2012An}
K.~K. Kim and M.~H. Kim, ``An enhanced anonymous authentication and key
  exchange scheme using smartcard,'' in \emph{Proc. of the 15th international
  conference on Information Security and Cryptology}, 2012, pp. 487--494.

\bibitem{Wang2015Preserving}
D.~Wang, N.~Wang, P.~Wang, and S.~Qing, ``Preserving privacy for free:
  Efficient and provably secure two-factor authentication scheme with user
  anonymity,'' \emph{Information Sciences}, vol. 321, no.~C, pp. 162--178,
  2015.

\bibitem{Li2009Key}
Y.~Li, Z.~Wu, and Q.~Liu, ``Key establishment and authentication scheme for
  heterogeneous integrated network based on cpk,'' \emph{Journal of Computer
  Applications}, vol.~29, no.~S2, pp. 72--75, 2009.

\bibitem{Chu2013An}
F.~Chu, R.~Zhang, R.~Ni, and W.~Dai, ``An improved identity authentication
  scheme for internet of things in heterogeneous networking environments,'' in
  \emph{Proc. of 2013 16th International Conference on Network-Based
  Information Systems(NBIS)}, 2014, pp. 589--593.

\bibitem{Amin2016Design}
R.~Amin, S.~K.~H. Islam, G.~P. Biswas, M.~K. Khan, L.~Lu, and N.~Kumar,
  ``Design of an anonymity-preserving three-factor authenticated key exchange
  protocol for wireless sensor networks,'' \emph{Computer Networks}, vol. 101,
  no.~C, pp. 42--62, 2016.

\bibitem{Arasteh2016A}
S.~Arasteh, S.~F. Aghili, and H.~Mala, ``A new lightweight authentication and
  key agreement protocol for internet of things,'' in \emph{Proc. of 2016 13th
  International Iranian Society of Cryptology Conference on Information
  Security and Cryptology (ISCISC)}, 2016, pp. 52--59.

\bibitem{Mahmood2017Lightweight}
Z.~Mahmood, H.~Ning, and A.~Ghafoor, ``Lightweight two-level session key
  management for end user authentication in internet of things,'' in
  \emph{Proc. of 2016 IEEE International Conference on Internet of Things
  (iThings) and IEEE Green Computing and Communications (GreenCom) and IEEE
  Cyber, Physical and Social Computing (CPSCom) and IEEE Smart Data
  (SmartData)}, 2016, pp. 323--327.

\bibitem{Sarvabhatla2014A}
M.~Sarvabhatla and C.~S. Vorugunti, ``A secure biometric-based user
  authentication scheme for heterogeneous wsn,'' in \emph{Proc. of 2014 Fourth
  International Conference of Emerging Applications of Information Technology},
  2014, pp. 367--372.

\bibitem{Iqbal2016A}
M.~A. Iqbal and M.~Bayoumi, ``A novel authentication and key agreement protocol
  for internet of things based resource-constrained body area sensors,'' in
  \emph{Proc. of 2016 IEEE 4th International Conference on Future Internet of
  Things and Cloud Workshops (FiCloudW)}, 2016, pp. 315--320.

\bibitem{Alkuhlani2017Lightweight}
A.~M.~I. Alkuhlani and S.~B. Thorat, ``Lightweight anonymity-preserving
  authentication and key agreement protocol for the internet of things
  environment,'' in \emph{Proc. of Smart Secure Systems - IoT and Analytics
  Perspective}, 2018, pp. 108--125.

\bibitem{Hou2008CPK}
H.~Hou, K.~Huang, and G.~Liu, ``Cpk and ecc-based authentication and key
  agreement scheme for heterogeneous wireless network,'' in \emph{Proc. of 2008
  International Conference on Computer Science and Software Engineering}, 2008,
  pp. 1015--1019.

\bibitem{Wang2017An}
C.~Wang, C.~Liu, S.~Niu, L.~Chen, and X.~Wang, ``An authenticated key agreement
  protocol for cross-domain based on heterogeneous signcryption scheme,'' in
  \emph{Proc. of 2017 13th International Wireless Communications and Mobile
  Computing Conference (IWCMC)}, 2017, pp. 723--728.

\bibitem{Luo2018Secure}
M.~Luo, Y.~Luo, Y.~Wan, and Z.~Wang, ``Secure and efficient access control
  scheme for wireless sensor networks in the cross-domain context of the iot,''
  \emph{Security and Communication Networks}, vol. 2018, pp. 1--10, 2018.

\end{thebibliography}

\end{document}